\documentclass[aps, twocolumn, pre, floatfix, superscriptaddress,longbibliography]{revtex4-2}

\usepackage{amsmath,amssymb,amsthm}
\usepackage{graphicx}

\usepackage{color}

\definecolor{darkblue}{rgb}{0,0,.65}
\definecolor{darkgreen}{rgb}{0.3,0.6,0.3}
\definecolor{cyan1}{rgb}{0.0, 0.6, 0.6}
\usepackage[%
pdfstartview=FitH,%
breaklinks=true,%
bookmarks=true,%
colorlinks=true,%
anchorcolor=black,%
citecolor=red,
filecolor=black,%
menucolor=black,%
urlcolor=darkblue,%
linkcolor=blue,%
]{hyperref}

\usepackage{cleveref}

\begin{document}

\title{Chaos in the three-site Bose-Hubbard model --- classical vs quantum}

\begin{abstract}

  We consider a quantum many-body system --- the Bose-Hubbard system on three sites --- which has a
  classical limit, and which is neither strongly chaotic nor integrable but rather shows a mixture
  of the two types of behavior.  We compare quantum measures of chaos (eigenvalue statistics and
  eigenvector structure) in the quantum system, with classical measures of chaos (Lyapunov
  exponents) in the corresponding classical system.  As a function of energy and interaction
  strength, we demonstrate a strong overall correspondence between the two cases.  In contrast to
  both strongly chaotic and integrable systems, the largest Lyapunov exponent is shown to be a
  multi-valued function of energy.  

\end{abstract}

\newcommand{\maynoothTP}{Department of Theoretical Physics, Maynooth University, Co.\ Kildare, Ireland}

\newcommand{\dresdenTP}{Institut f\"{u}r Theoretische Physik, Technische Universit\"{a}t Dresden, D-01062 Dresden, Germany}

\author{Goran Nakerst}
\affiliation{\maynoothTP}
\affiliation{\dresdenTP}

\author{Masudul Haque}
\affiliation{\maynoothTP}
\affiliation{\dresdenTP}
\affiliation{Max-Planck-Institut f\"{u}r Physik komplexer Systeme, D-01187 Dresden, Germany}

\maketitle

\section{Introduction and Overview}

How to translate classical chaos to quantum systems has been studied since the very beginning of
quantum mechanics \cite{Einstein_VerhDeuPhy_1917_Quantensatz}.  Classical chaos is the sensitivity of
dynamics to initial perturbations \cite{Schuster_2006}, while quantum chaos manifests itself in
various quantum properties, such as in level spacing distributions
\cite{Wigner_1957, Haake_book_quantum_chaos, Stockmann_2000_quantum_chaos}.
Level spacings of quantum models with an integrable classical limit typically follow Poisson's law
\cite{Berry_Tabor_PreRoySocLon1977}, while level spacings of models with a classically chaotic limit
typically obey Wigner distributions \cite{Bohigas_Giannoni_Schmidt_PRL1984,
  Bohigas_Giannoni_Schmidt_JdPL1984} typical of random matrices.  In this work we will address the
question how classical chaos relates to quantum notions of chaos in a many-body system that is
neither integrable nor strongly chaotic, but rather shows a mixture of both behaviors, i.e., is a
``mixed'' system.

Regular and chaotic motion coexist in the phase space of mixed classical Hamiltonian systems, and corresponding quantum systems show a combination of chaotic and non-chaotic features. For
Hamiltonian systems with a few degrees of freedom, mixed behavior is considered generic
\cite{Feingold_Peres_PhysicaD1983_coupled_rotators, Feingold_Moiseyev_Peres_PRA1984, Wintgen_Fridrich_PRA1987, 
  Bohigas_Tomsovic_Ullmo_PhysRep1993, Dullin_Richter_Wittek_Chaos1996, Robb_Reichl_PRE1998_twospin,
  Ketzmerick_Hufnagel_Steinbach_Weiss_PRL2000, Dullin_Baecker_Nonlinearity2001,
  Emerson_Ballentine_PRA2001_twospin, Baecker_EPJST2007_randomwaves,
  Fan_Gnutzmann_Liang_PRE2017_FeingoldPeres, Batistic_Lozej_Robnik_PRE2019}.  In contrast, mixed systems are less commonly
considered in many-body systems.  In particular, in the thermodynamic limit, proximity to
integrability is expected to have negligible effects and systems are expected to be driven to be
chaotic/ergodic by infinitesimal perturbations away from integrable points.  In this work, we will
consider the classical limit instead of the thermodynamic limit, by considering a system of $N$
bosons and taking the $N\to\infty$ limit on a fixed lattice geometry.  For a small number of lattice
sites, this can lead to mixed behavior both for the quantum system and for the classical limit.

To connect the classical and quantum worlds we will refine the classical phase space into energy
manifolds and compare with eigenvalues/eigenstates of the quantum Hamiltonian in the corresponding
energy ranges.

The sensitivity to initial conditions of the classical motion will be measured by the largest
Lyapunov exponent $\lambda_{\rm max}$.  It is generically not possible to calculate the largest
Lyapunov exponents analytically or exactly; we will therefore estimate them numerically by
integrating classical motion up to a finite time, the finite-time Lyapunov exponents (FTLEs).  We
will use the terms ``Lyapunov exponents'' and ``FTLEs'' interchangeably --- it is to be understood
that all presented data for $\lambda_{\rm max}$ are best available numerical estimates and that
analytically exact values are generally not available.

The model we consider is the celebrated Bose-Hubbard model. It attains a classical limit for fixed
number of sites $L$ and increasing particle number $N$. In this limit it is equivalent to the
well-known discrete non-linear Schr\"odinger equation (DNLS)
\cite{Trombettoni_Smerzi_PRL2001_solitonsbreathers, Smerzi_Trombettoni_Kevrekidis_Bishop_PRL2002,
  Schmerzi_Trombettoni_2003_tight_binding_BEC, book_Kevrekidis_springer2009}, which is a classical
Hamiltonian system.  The DNLS can be regarded as a mean-field approximation or as a classical limit
of the Bose-Hubbard model.  As a classical limit, it is widely used as the basis for semi-classical
studies of the Bose-Hubbard model
\cite{ Polkovnikov_semiclassics_PRA2003_TWA, Mahmud_Reinhardt_PRA2005_semiclassical_BHdimer,
  Hiller_Kottos_Geisel_PRA2006, Mossmann_Jung_PRA2006_semiclassical_BHtrimer,
  Graefe_Korsch_PRA2007_semiclassical_BHdimer, Trimborn_Witthaut_Korsch_PRA2009,
  Cassidy_PRL2009_BHclassical, Polkovnikov_AnnPhys2010_PhaseSpace, Trippenbach_PRA2011_BHdimer,
  Simon_Strunz_PRA2012_BHDimer, Simon_Strunz_PRA2014_BHDimer_Trimer,
  Veksler_Fishman_NJP2015_BHDimer_semiclassical,
  Engl_Urbina_Richter_PRE2015_semiclassical_traceformula,
  Engl_Urbina_Richter_PhilTrans2016_semiclassical, Grossmann_Strunz_JPA2016_semiclassical,
  Dubertrand_Mueller_NJP2016_semiclassics_spectralstat, Kidd_arxiv2017_BHclassical,
  Tomsovic_Schlagheck_Richter_PRA2018_postEhrenfest, Tomsovic_PRE2018_saddle_BH,
  Rammensee_Urbina_Richter_PRL2018_semiclassicOTOCs, Corney_PRA2019_BHdimer,
  Schlagheck_Ullmo_Urbina_Richter_Tomsovic_PRL2019}.
The behavior of the quantum model has been compared to that of the DNLS
\cite{%
  Milburn_Munro_PRA2000_BHtrimer, Weiss_Teichmann_PRL2008_BHdimer_quantumclassical,
  Viscondi_Furuya__JPhysA2011_BHtrimer, Gertjerenken_Weiss_PRA2013_BHdimer_quantumclassical,
  Kolovsky_IJMPB2016, Heinisch_Holthaus_ZNforschungA2016, Rautenberg_Gaertner_PRA2020,
  Pausch_Buchleitner_PRL2021_BHchaos, Pausch_Buchleitner_NJP2021_BHchaos,
  Castro_Santos_Hirsch_OpenAcc2021}.
When considering the classical limit, it is convenient to parametrize the interaction by
$\Lambda = UN$, where $U$ is the on-site interaction and $N$ is the number of particles. In the
limits $\Lambda \to 0$ and $\Lambda \to \infty$ it can be analytically solved and is therefore
integrable. In the special case of $L=2$ sites the model is integrable for all $\Lambda$.   For
finite $\Lambda$ the Bose-Hubbard model on three or more sites is known to be non-integrable
\cite{Kolovsky_Buchleitner_EPL2004, Kollath_JoSM2010}.
Despite being non-integrable, for $L=3$ and to some extent for $L=4$ the system is not strongly
chaotic but rather highly mixed
\cite{Mossmann_Jung_PRA2006_semiclassical_BHtrimer, Hiller_Kottos_Geisel_BHtrimer_PRA2009,
  Viscondi_Furuya__JPhysA2011_BHtrimer, Geva_Vardi_Cohen_PRA2014_BHtrimer,
  Vardi_RomRepPhy2015_BHtrimer, Han_Wu_PRA2016_BHtrimer_ehrenfest, Buerkle_Anglin_4siteBH_PRA2019,
  Kolovsky_russians_AIPConf2020, Rautenberg_Gaertner_PRA2020, Nakerst_Haque_PRE2021,
  Wittmann_Santos_arxiv2021}. %
We will mainly restrict our analysis to $L=3$ sites.  (In the appendix, for comparison we
present some classical results for $L=4$ and $L=7$ sites.)

\begin{figure*}
\begin{center}
\includegraphics[width=\textwidth]{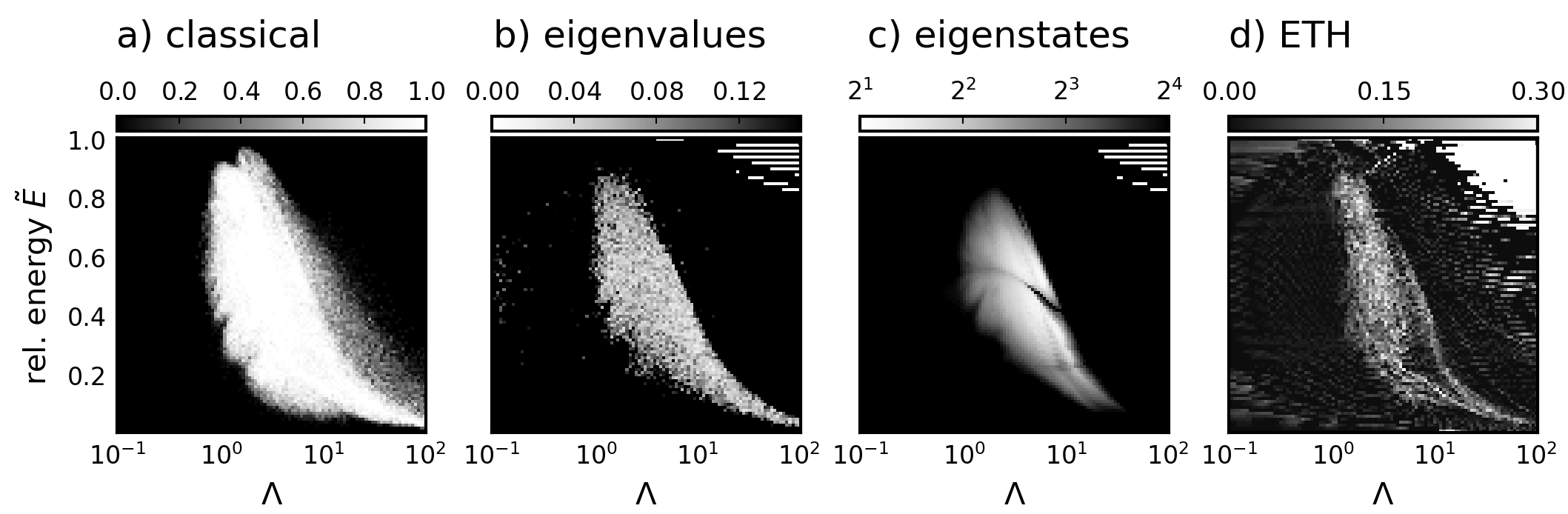}
\caption{\label{fig:chaos_2d}%
Heatmap of classical chaos indicator \textbf{a)} and quantum chaos
indicators \textbf{b)}, \textbf{c)} and \textbf{d)}. Relative energy 0 corresponds
to the minimal (ground state) energy while 1 corresponds to the maximal energy and $\Lambda$ is the onsite interaction strength. The lighter the color the more chaotic.
\textbf{a)} Fraction of states with positive largest Lyapunov exponent
\textbf{b)} Kullback-Leibler divergence of the distribution of level ratios from the distribution of level ratios of Gaussian orthogonal matrices; capped at 0.15.
\textbf{c)} Excess kurtosis of eigenstates; capped at $2^4$.
\textbf{d)} Exponent of the exponential decay of ETH fluctuations with Hilbert space size of normalized operators, clipped between 0 and 0.3.
}
\end{center}
\end{figure*}

We will compare the FTLEs for the classical system to several chaos indicators on the quantum side
--- level statistics, eigenstate statistics, and the scaling of fluctuations of eigenstate
expectation values (EEVs).
Figure \ref{fig:chaos_2d} provides an overview of the results.
Here we show chaoticity as a function of interaction parameter $\Lambda$ and relative energy.  Chaos
is visualized as greyscale heatmaps, where the intensity indicates how chaotic that region is ---
the lighter the more chaotic.

Figure \ref{fig:chaos_2d} a) shows chaos of the \textit{classical} Bose-Hubbard model, while b), c)
and d) show chaos measures of the \textit{quantum} system.  In a) we show the fraction of positive
FTLEs of the classical model. We consider a FTLE as positive if it is greater than $10^{-4}$, and
zero otherwise. In b) we show the deviation of level statistics of the quantum model from Wigner's
GOE distribution measured by the Kullback-Leibler divergence. In c) we show how much eigenstates of
the quantum model deviate from Gaussian states via the kurtosis.  The kurtosis obtained from the
eigenstate coefficients in two different bases are combined --- the larger of the two is used at
every point of the heatmap.  In d) we show the exponent in the size-dependence of the fluctuations
of eigenstate expectation values (EEV's).  This is motivated by the scaling of the eigenstate
thermalization hypothesis (ETH) and may be described as the size dependence of ETH fluctuations in
approaching the classical limit \cite{Nakerst_Haque_PRE2021}.  The data in panels b), c) are for
$N=150$ bosons, while the exponents in d) are obtained by fitting EEV fluctuations between $N=90$
and $N=170$.  Overall, we have found these quantum results to be broadly independent of $N$.

Figure \ref{fig:chaos_2d} a) reveals features of the classical phase space, i.e., the phase space of
the three-site discrete nonlinear Schr\"odinger equation. For $\Lambda\lesssim1$ all Lyapunov
exponents are close to 0. For $\Lambda>1$ regions with a non-zero fraction of positive Lyapunov
exponents emerge at intermediate energies.  At $\Lambda \approx 3$ there are positive largest FTLEs
at most energies, except for smallest and largest energies. For $\Lambda > 3$ the region of non-zero
fractions of positive Lyapunov exponents shrinks and shifts to lower energies, where it survives
even for the largest $\Lambda=100$ we investigated.  These results highlight the mixed nature of the
classical phase space.  In particular, zero and non-zero Lyapunov exponents exist at the same energy
for the same $\Lambda$.  This will be further elaborated in section \ref{sec:Lyapunov_exponents} and
figure \ref{fig:lyapunov_vs_energy}.

The same shape of the heatmap in figure \ref{fig:chaos_2d} a) is observed in b) -- d) as well. The
white bars at the top right of the quantum plots do not show chaotic regions; these are finite size
defects (gaps in the spectra which are larger than the energy windows used to compile the heatmaps).
The exact measures used in these panels and the subtleties encountered for the quantum cases will be
detailed in Sections \ref{sec:eigenvalues}, \ref{sec:eigenstates} and \ref{sec:eth}, which focus
respectively on level statistics, panel b), on eigenstate amplitude statistics, panel c), and on EEV
scaling, panel d).

The overall visual agreement between classical chaos regions, panel a), and quantum chaos regions,
panels b)--d), is striking.  Chaotic energy regions of the classical phase space correspond
generally to chaotic regions of the spectrum of the quantum Hamiltonian.
Even fine structures in the heatmaps show some agreement. For $1 < \Lambda< 3$ small bulbs appear at
the chaotic-regular boundary in the classical spectrum a), which can be recognized in the level
statistics b) as well as in the kurtosis of eigenstates c).  We conclude that overall there is a
close correspondence of chaotic and non-chaotic regions of the classical model and the quantum
model.  There are, of course, some discrepancies, also among the various quantum measures, and
various artifacts due to the particular measures used.  These issues will be discussed in the body
of this paper.

In section \ref{sec:model} we introduce the version of the quantum Bose-Hubbard model we use, and
its classical limit.  Lyapunov exponents of the classical model are analyzed in section
\ref{sec:Lyapunov_exponents}, where the data of figure \ref{fig:chaos_2d}a) is explained, and other
ways of using the FTLEs to demarcate chaotic and non-chaotic regions are explored.  In section
\ref{sec:eigenvalues} we investigate the eigenvalues of the quantum model leading to the results
shown in figure \ref{fig:chaos_2d}b).  In section \ref{sec:eigenstates} eigenstates are compared to
Gaussian states and the numerical derivation of figure \ref{fig:chaos_2d}c) is explained.  In
section \ref{sec:eth}, we describe quantifying chaos using EEV scaling exponents, and explain how
figure \ref{fig:chaos_2d}d) is obtained.  We end the main text with some discussion in section
\ref{sec:discussion}.  In the Appendices, we provide an outline of how to calculate Lyapunov
exponents, and we also show and briefly discuss Lyapunov exponents for $L=4$ and $L=7$ sites.

\section{Model and parametrizations}\label{sec:model}

\subsection{Quantum model and classical limit}

We will investigate Bose-Hubbard systems restricted to open-boundary chains of length $L$, with
nearest neighbor hoppings and on-site interactions.  The quantum Hamiltonian is
\begin{equation}\label{eq:Bose_Hubbard_quantum_H}
	H = -\frac{1}{2} \sum_{\langle j,l\rangle} J_{j,l} a_j^\dagger a_l + \frac{U}{2}\sum_{j=1}^L n_j(n_j-1),
\end{equation}
where $\langle j,l \rangle$ denotes summation over adjacent sites ($l=j\pm1$), $a_j^\dagger$ and
$a_j$ are the bosonic creation and annihilation operators for the $j$-th site and
$n_j=a_j^{\dagger}a_j$ is the corresponding occupation number operator. $J_{j,l} = J_{l,j}$ is the
symmetric tunneling coefficient and $U$ is the two-particle on-site interaction strength.  We choose
$J_{1,2}=1.5$ and $J_{j,l}=1$ for $j,l\ge 2$ to avoid reflection symmetry.
The number of bosons $N$ is conserved by the Hamiltonian.  We introduce the tuning parameter
$\Lambda=UN$.
The Hilbert space dimension $D$ of the quantum Hamiltonian is
$D=\binom{N+L-1}{L-1}$.  For constant $L$, this grows with  boson
number $N$ as $D \sim N^{L-1}$.

The Bose-Hubbard model has a classical limit for $N\to\infty$ while keeping the number of sites $L$
fixed. This limit can be taken by renormalizing the creation and annihilation operators via
$a \to a/\sqrt{N} = \bar{a}$, so
$[\bar{a}^\dagger_j, \bar{a}_j] = 1/N = \hbar_{eff}  \to 0$ for $N\to \infty$, where we let $\hbar=1$. The renormalized Hamiltonian
is then given by
\begin{equation}
\bar{H} = H/N = -\frac{1}{2} \sum_{\langle j,l\rangle} J_{j,l} \bar{a}_j^\dagger \bar{a}_l + \frac{\Lambda}{2} \sum_j \bar{n}_j(\bar{n}_j - 1/N),
\end{equation}
where $\bar{n}=\bar{a}^\dagger \bar{a}$. The corresponding classical Hamiltonian $\mathcal{H}$ in
the large $N$-limit is then obtained by replacing the operators $\bar{a}$,  $\bar{a}^{\dagger}$ by complex
numbers $\psi$, $\psi^*$:
\begin{equation}\label{eq:Bose_Hubbard_classical_H}
\mathcal{H} = - \frac{1}{2} \sum_{\langle j,l\rangle} J_{j,l} \psi_j^* \psi_l +
\frac{\Lambda}{2}\sum_j |\psi_j|^4. 
\end{equation}
Conservation of the total number of particles $N$ in the quantum systems enforces
\begin{equation} \label{eq:norm_conservation}
	\sum_j |\psi_j|^2 = 1,
\end{equation}
so the phase space of the classical model is restricted to the real hyper-sphere $S^{2L-1} \subset \mathbb{R}^{2L}$. 

The dynamics of the classical Bose-Hubbard Hamiltonian are given by Hamilton's equations of motion
\begin{equation}\label{eq:Bose_Hubbard_eom}
	i\frac{\partial}{\partial t} \psi_j 
	= \frac{\partial \mathcal{H}}{\partial \psi_j^*}
	= -\frac{1}{2} \sum_{\langle j,l\rangle} J_{j,l} \psi_l + \Lambda  |\psi_j|^2\psi_j,
\end{equation}
which is also known as the discrete nonlinear Schr\"odinger equation, or the discrete
Gross-Pitaevskii equation.

Identifying the complex plane $\mathbb{C}$ with the real plane $\mathbb{R}^2$ one can rewrite the
$L$ complex equations in Eq.~\eqref{eq:Bose_Hubbard_eom} as $2L$ real equations. For computational
reasons we chose Cartesian coordinates and let $x=\operatorname{Re} \psi$ and
$y=\operatorname{Im} \psi$. Then Hamilton's equation of motion read
\begin{equation}
	\frac{\partial \mathcal{H}}{\partial x_j}
	= - \sum_{\langle j,k\rangle} J_{j,k} x_k + 2\Lambda \sum_j (x_j^2 + y_j^2) x_j
\end{equation}
and 
\begin{equation}
	\frac{\partial \mathcal{H}}{\partial y_j}
	= - \sum_{\langle j,k\rangle} J_{j,k} y_k + 2\Lambda \sum_j (x_j^2 + y_j^2) y_j.
\end{equation}

In the limits $\Lambda\to 0$ and $\Lambda\to\infty$, both the quantum
and the classical systems are integrable. If $\Lambda=0$ then
Eq.~\eqref{eq:Bose_Hubbard_quantum_H} and
Eq.~\eqref{eq:Bose_Hubbard_classical_H} reduce to free particle
Hamiltonians.

If $\Lambda\to\infty$ then one effectively can neglect $J$, so that Eq.~\eqref{eq:Bose_Hubbard_quantum_H}
is diagonal in the computational basis of mutual eigenstates of $n_j$ and the equations of motion Eq.~\eqref{eq:Bose_Hubbard_eom} are decoupled. 

In the remainder of the paper we mainly focus on the Bose-Hubbard model on $L=3$ sites, so as to
focus on a system with very mixed behavior.  Computationally, since the Hilbert space size $D$ grows
only quadratically in $N$ for $L=3$, we are able to consider the quantum Hamiltonian far into the
classical limit (small $\hbar_{eff} = 1/N$) while keeping the Hilbert space size accessible for
numerical diagonalization.

\subsection{Relative energy and energy intervals \label{subsec_energy_intervals}}

Our classical-quantum comparison is energy-resolved.
For each $\Lambda$, we will compare the degree of chaos in individual energy regions of the
classical system with the degree of chaos in corresponding energy regions of the quantum system.

Numerically, for each interaction $\Lambda$ the possible energies are divided into 100 evenly spaced
energy intervals.  We also rescale and shift the energy for each $\Lambda$ to define the relative
energy
\begin{equation}
\tilde{E}= \frac{E - E_{\min}}{E_{\max}- E_{\min}}
\end{equation}
which takes values in the range $[0,1]$.  For the classical system, $E_{\min}$ and $E_{\max}$ are
the lowest and highest possible classical energies.  For the quantum system, they are respectively
the lowest eigenenergy (ground state energy) and the highest eigenenergy.  Each energy interval
corresponds to an interval of $\tilde{E}$ having width $0.01$.  When we refer to the interval at
relative energy $\tilde{E}$, we mean the interval $[\tilde{E}-0.01,\tilde{E}]$.

For the classical calculation (figure \ref{fig:chaos_2d}a, section \ref{sec:Lyapunov_exponents}),
Lyapunov exponents are collected for phase space points whose energy is in the desired interval.
For the quantum eigenvalue statistics (figure \ref{fig:chaos_2d}b, section \ref{sec:eigenvalues}),
the spacing between eigenvalues within the desired interval is analyzed.  For quantum measures based
on eigenstates (figure \ref{fig:chaos_2d}c and \ref{fig:chaos_2d}d, sections \ref{sec:eigenstates}
and \ref{sec:eth}), all eigenstates whose eigenvalues lie in the interval are considered.

\begin{figure*} 
\begin{center} 
\includegraphics[width=\textwidth]{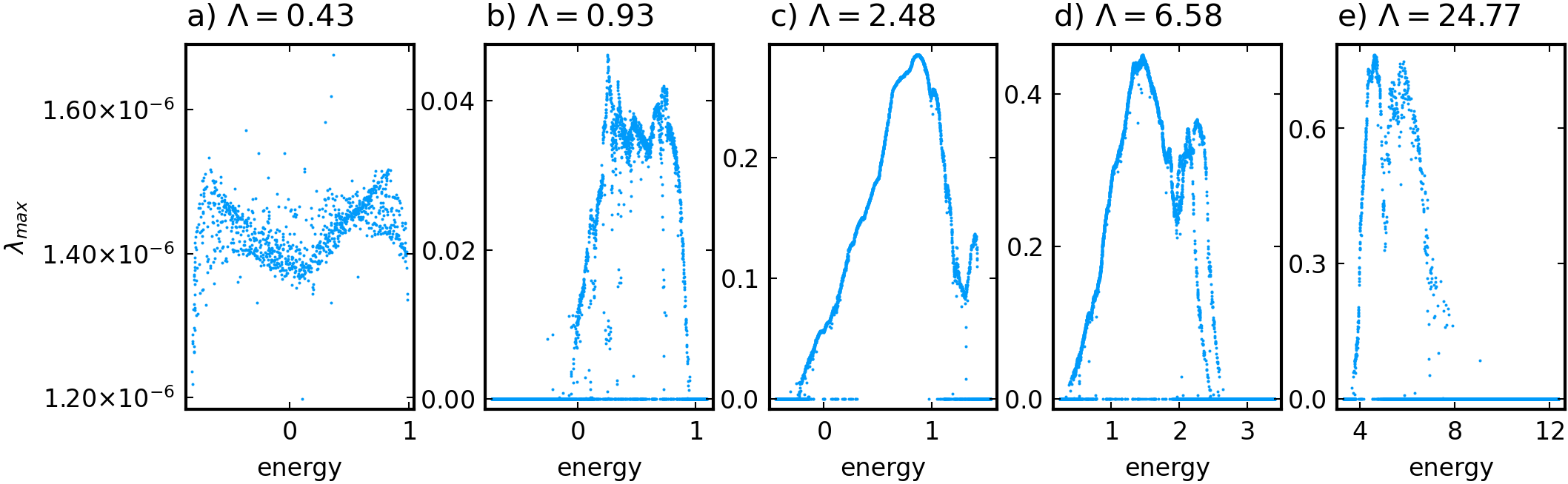} 
\caption{\label{fig:lyapunov_vs_energy} 
FTLE estimates for largest Lyapunov exponent $\lambda_{\text{max}}$ for the classical limit, Eq.\
\eqref{eq:Bose_Hubbard_eom}.  The numerical estimates for $\lambda_{\text{max}}$ are plotted against
energy, for several different values of the interaction parameter $\Lambda$. 
} 
\end{center}
\end{figure*}

\section{Classical Lyapunov exponents}\label{sec:Lyapunov_exponents}

In this section we describe the largest Lyapunov exponents, $\lambda_{\rm max}$, of the classical
model, Eq.\ \eqref{eq:Bose_Hubbard_eom}.  Some background on Lyapunov exponents and their numerical
calculation (calculating FTLE's) is provided in Appendix
\ref{sec:appendix_numerical_Lyapunov_exponents}.

In ergodic Hamiltonian systems, $\lambda_{\rm max}$ depends solely on the single conserved quantity,
the energy.  We will show that the Bose-Hubbard system on three sites is not an ergodic system in
this sense --- there exist states with the same energy but different largest Lyapunov exponents.

\subsection{Preliminaries}

For the $L$-site system, because there are $2L$ real equations of motion, we have $2L$ different
Lyapunov exponents $\lambda_{\text{max}} = \lambda_1 \ge \dots \ge \lambda_{2L}$.  These exponents
are arranged in $L$ pairs of equal magnitude and opposite sign.  Two pairs are zero because of the
conservation of energy and the conservation of norm, Eq.\ \eqref{eq:norm_conservation}.  Thus at
most $L-2$ exponents can be positive.  For $L=3$, which we will focus on, there is at most one
positive Lyapunov exponent.  This is $\lambda_{\text{max}}$.

The classical phase space is continuous, so a numerical calculation of the Lyapunov exponents for
all states is not possible.  One could sample states according to a uniform distribution on the
phase space.  Due to the conservation of total norm \eqref{eq:norm_conservation}, the classical
phase space is restricted to the sphere $S^5$ in $\mathbb{R}^6$.  Choosing the $2L$ components of
the state from a Gaussian distribution, and then normalizing the resulting state, amounts to
sampling uniformly on $S^5$.  Unfortunately, sampling uniformly on $S^5$ can lead to a dearth of
samples for small energies and large energies.  To sample uniformly in energy we divide the energy
spectrum into 100 evenly spaced intervals and sample states uniformly within these energy intervals
by the rejection method.  We sample states uniformly on $S^5$ and reject samples unless the energy
of the state lies in the desired energy interval. In this way for each interaction $\Lambda$ we
sample up to $10^4$ states uniformly distributed in energy.

Obtaining good estimates of Lyapunov exponents is numerically challenging for imperfectly chaotic
systems.  The system needs to be evolved for long times.  The FTLE's presented in this work are
obtained by evolving up to one million time units.

\subsection{Numerical observations}

In figure \ref{fig:lyapunov_vs_energy} we plot FTLE's of sampled states against the energy of these
states, for several interaction parameters $\Lambda$.  Only estimates of the largest Lyapunov
exponent $\lambda_{\rm max}$ are presented --- the other LE's are either zero or the negative of
$\lambda_{\rm max}$.

For $\Lambda=0$, the  model is integrable and hence $\lambda_{\text{max}}=0$.

Figure \ref{fig:lyapunov_vs_energy}a) shows the numerical estimates for $\lambda_{\text{max}}$ for
non-zero but still small $\Lambda$ ($\Lambda\approx 0.43$).  The numerical estimates for all six
Lyapunov exponents have the same order of magnitude, $10^{-6}$.  This implies that
$\lambda_{\text{max}}$ is either zero or vanishingly small up to some finite value of the
interaction.

For larger $\Lambda$, panels b)-e), we find cases of $\lambda_{\text{max}}$ being unambiguously
zero, together with cases of the FTLE being smaller than the cutoff $10^{-4}$, which we interpret as
$\lambda_{\text{max}}$ being zero.  In each of these panels, there are low-energy and high-energy
regimes where there are only zero $\lambda_{\text{max}}$, and a central energy regime with nonzero
positive $\lambda_{\text{max}}$.  For smaller $\Lambda$, the positive-$\lambda_{\text{max}}$
behavior is concentrated at higher energies (there is an extended $\lambda_{\text{max}}=0$ range of
low energies), panel b).  For large $\Lambda$, the converse is true: $\lambda_{\text{max}}>0$ is
seen at lower energies, panel d),e).

In general, when there are nonzero exponents, they coexist with zero exponents at the same energy,
i.e., the $\lambda_{\text{max}}$ vs energy function is multi-valued.  The only exception is in the
intermediate-interaction panel c), $\Lambda\approx 2.48$, for which an energy window with a single
nonzero branch is seen.  In fact, for any $\Lambda \gtrapprox 1$, there appears always to be some
energy window where $\lambda_{\text{max}}$ is multi-valued --- we did not see any exceptions.

The coexistence of zero and nonzero $\lambda_{\text{max}}$ is a peculiar manifestation of the mixed
nature of the system.  This is in contrast to integrable systems (for which $\lambda_{\text{max}}$
is always zero except a measure zero set) and to strongly chaotic or ergodic systems (for which
$\lambda_{\text{max}}$ is always positive except a measure zero set).  To highlight this contrast,
we give examples of systems with stronger chaos, the same Hamiltonian on $L=4$ sites and $L=7$ sites, in
the appendix, figure \ref{fig:lyapunov_vs_energy_large_L}.

In an ergodic system the largest Lyapunov exponent is a smooth single valued function of energy.  We
showed that $\lambda_{\rm max}$ is not a single valued function, but rather often has two branches.
One can ask whether each branch is smooth.  There are some noisy features in the plots, especially
in panels a), b) and e).  Presumably, these are finite-time effects, and each branch would resolve
into smooth lines if we could integrate up to infinite times.  While this conjecture could not be
verified conclusively, we observed that integrating up to longer times generally reduces the noisy
aspect.

In one case, panel d), $\lambda_{\rm max}$ even appears to have three branches (one zero and two
nonzero).  We have not seen any indication that this is a finite-time effect, although we cannot
rule it out.  The data suggests that the mixed nature of the system even allows for three
$\lambda_{\rm max}$ values.  Apparently, the same fixed-energy region of phase space can consist of
a regular (non-chaotic) sub-manifold as well as two different sub-manifolds with different nonzero
$\lambda_{\rm max}$!

In figure \ref{fig:chaos_2d} a) we used as an indicator of chaos the fraction of $\lambda_{\rm max}$
which are nonzero; the same measure has been used in Ref.\
\cite{Carlos_Santos_Hirsch_PRL2019_LyapunovOTOCDickemodel}.

\subsection{Using the magnitude of Lyapunov exponents} \label{subsec:magnitude_Lyapunovs}

\begin{figure}
\begin{center}
\includegraphics[width=\columnwidth]{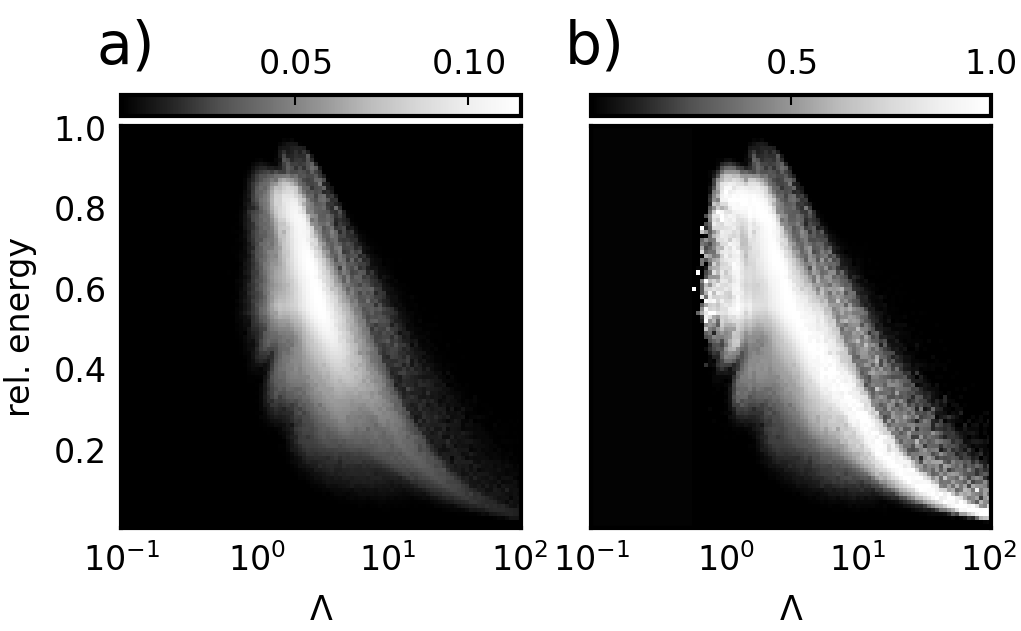}
\caption{\label{fig:chaos2d_lyapunov} Average largest Lyapunov exponent $\bar{\lambda}_{\rm max}$
  per relative energy interval renormalized by \textbf{a)} $\beta_\Lambda$ given by
  Eq.~\eqref{eq:beta_Lambda} and \textbf{b)} $\gamma_\Lambda$ given by Eq.~\eqref{eq:gamma_Lambda}.} 
\end{center}
\end{figure}

The procedure of using the fraction of non-zero $\lambda_{\rm max}$'s to characterize chaos neglects
the magnitudes of $\lambda_{\rm max}$ altogether.  One could also make use of the magnitude as a
chaos indicator.  This raises the issue of comparing values of $\lambda_{\rm max}$ for different
systems (systems with different interactions $\Lambda$).  We consider two ways of rescaling the
$\lambda_{\rm max}$ values; the resulting heatmaps in figure \ref{fig:chaos2d_lyapunov} show
reasonable agreement with that in figure \ref{fig:chaos_2d}a).

The magnitude of $\lambda_{\text{max}}$ depends on the timescales of the dynamics of the
system.
From Eq.~\eqref{eq:Bose_Hubbard_eom} one could expect that the dominant timescale will be given by
the inverse of the maximum of the Hamiltonian parameters, $J$ and $\Lambda$. We fixed $J_{12}=1.5$
and $J_{23}=1$ throughout the paper so $\max(J,\Lambda)=\max(1.5, \Lambda)$. In figure
\ref{fig:chaos2d_lyapunov} a) we show the average largest Lyapunov exponent
$\overline{\lambda}_{\text{max}}$ per energy interval, rescaled by
\begin{equation}\label{eq:beta_Lambda}
	\beta_\Lambda = \max(1.5,\Lambda).
\end{equation}
The resulting heatmap in figure \ref{fig:chaos2d_lyapunov}a), by construction, shows appreciable
chaos in the same region as in figure \ref{fig:chaos_2d}a).  But figure \ref{fig:chaos2d_lyapunov}a)
shows more detail as it encapsulates the information about the magnitude of $\lambda_{\text{max}}$
as well. We observe the highest intensities in the mid of the spectrum for $1<\Lambda<10$. From
there it falls of in all directions.  At the top end of figure \ref{fig:chaos2d_lyapunov}a) we
observe a dip in intensity and a sudden increase again, before $\bar{\lambda}_{\rm max}$ becomes
zero. These reflect the dips seen in figure \ref{fig:lyapunov_vs_energy} c) and d).

Another approach is to rescale all largest Lyapunov exponents in a system with fixed interaction
$\Lambda$ by the maximal largest Lyapunov exponent $\lambda_{\text{max}}$ in that specific system. A
problem occurs when all largest Lyapunov exponents are close to zero, as for Bose-Hubbard systems
with $\Lambda \ll 1$.  In these systems there is simply no appreciable positive
$\lambda_{\text{max}}$.  Therefore we choose the cutoff $10^{-4}$ by which all Lyapunov exponents
are minimally divided. The rescaling parameter is
\begin{equation}\label{eq:gamma_Lambda}
	\gamma_\Lambda = \max(10^{-4}, \max_{\psi} \lambda_{\text{max}}(\psi) ),
\end{equation}
where the maximum runs over all states $\psi$ in the phase space and $\lambda_{\text{max}}(\psi)$
denotes the corresponding largest Lyapunov exponent.  A heatmap of the average largest Lyapunov
exponent $\bar{\lambda}_{\rm max}$ with this rescaling is shown in figure \ref{fig:chaos2d_lyapunov}b).

The overall features are the same as in panel \ref{fig:chaos2d_lyapunov}a).  There are some
artifacts at the boundary between chaotic and non-chaotic regions, around $\Lambda\approx 0.7$ in
panel b), presumably because of numerical uncertainties when $\lambda_{\text{max}}$ is around
$10^{-4}$.  The intensity of the heatmap does not decrease with $\Lambda$ beyond
$\Lambda\approx 10$, unlike panel a) where this decrease is built into the scaling function
$\beta_\Lambda$.

\section{Eigenvalue statistics} \label{sec:eigenvalues}

In this section we will compare the distribution of spacing ratios of energy levels of the
Bose-Hubbard trimer to the level ratio distribution of Gaussian orthogonal random matrices and
Poisson level ratios. We will compare the first moment and the whole distribution and describe how
we numerically obtain the results leading to figure \ref{fig:chaos_2d}b).

\subsection{Definitions \& Background}

For ordered energy levels $E_\alpha < E_{\alpha+1}$ of a Hamiltonian
we denote energy level spacings by $s_\alpha = E_{\alpha+1} -
E_\alpha$. Instead of examining the distribution of  $s_\alpha$
itself, it has become common to investigate instead the  distribution of spacing ratios \cite{Oganesyan_Huse_PRB_2007, Atas_Bogomolny_Roux_PRL_2013}
\begin{equation}
	r_\alpha = \frac{s_{\alpha+1}}{s_\alpha} \quad \text{and} \quad \tilde{r}_\alpha =
	\min\left(r_\alpha, \frac{1}{r_\alpha}\right).
\end{equation}
Studying the ratio distribution bypasses the need to unfold the energy
spectrum (to account for the density of states).  The quantity
$\tilde{r}_\alpha$ has the additional advantage that it has bounded
support, $\tilde{r}_\alpha\in[0,1]$.  In the following, when we refer
to level spacing ratios, we mean $\tilde{r}_\alpha$ (and not $r_\alpha$).

The level spacing ratio distribution for matrices in the Gaussian Orthogonal Ensemble (GOE) is given
approximately by 
\begin{equation}\label{eq:GOE_ratio_density}
	p_{\rm GOE}(\tilde{r}) = \frac{2}{Z_{GOE}} \frac{\tilde{r}+\tilde{r}^2}{(1+\tilde{r}+\tilde{r}^2)^{5/2}},
\end{equation}
where the normalization constant is  $Z_{GOE} = 8/27$.  This is the expected spectral behavior of  highly
chaotic systems.  For uncorrelated (Poisson) eigenvalues, the ratio distribution is 
\begin{equation}\label{eq:Poisson_ratio_density}
	p_{\rm Poi}(\tilde{r}) = \frac{2}{(\tilde{r}+1)^2}. 
\end{equation}
In both cases, the distribution  is understood to vanish outside $[0,1]$.  GOE spectra have level
repulsion, $p_{\rm GOE}(0)=0$, while Poisson spectra do not, $p_{\rm Poi}(0)\neq0$.

It is usual to compare level spacing distributions over the complete spectrum (sometimes omitting
spectral edges) with standard distributions like GOE or Poisson.  In this work, we would like to
characterize the degree chaos at different energies; hence we compare the distributions obtained
from the energy levels within each of the energy intervals described in Subsection
\ref{subsec_energy_intervals}.  Such energy-resolved comparisons of level statistics have appeared
e.g., in \cite{Luitz_Laflorencie_Alet_PRB2015, Pausch_Buchleitner_PRL2021_BHchaos,
  Nakerst_Haque_PRE2021}.

To compare with GOE or Poisson distributions, we use a common measure
of the difference between two distributions, namely the
Kullback-Leibler (KL) divergence \cite{Kullback_Leibler_1951}.  The KL
divergence between an observed distribution $P(z)$ and a reference
distribution  $Q(z)$ is
\begin{equation}\label{eqn_kld}
  D_{\mathrm{KL}}(P|Q) = \int_{-\infty}^\infty P(z)\log\frac{P(z)}{Q(z)}dz. 
\end{equation}
This quantity vanishes if $P(z)$ is identical to $Q(z)$.  Generally, a
larger KL divergence indicates stronger deviation of $P(z)$ from $Q(z)$.  In
this work, $P(z)$ will be the ratio distributions obtained from the
Bose-Hubbard energy levels within each energy interval.  We will use
either the GOE or the Poisson distribution, Eq.\
\eqref{eq:GOE_ratio_density} or \eqref{eq:Poisson_ratio_density}, as
the reference $Q(z)$.

\subsection{Numerical Observations --- entire distribution and KL divergence}

\begin{figure}
\begin{center}
\includegraphics[width=\columnwidth]{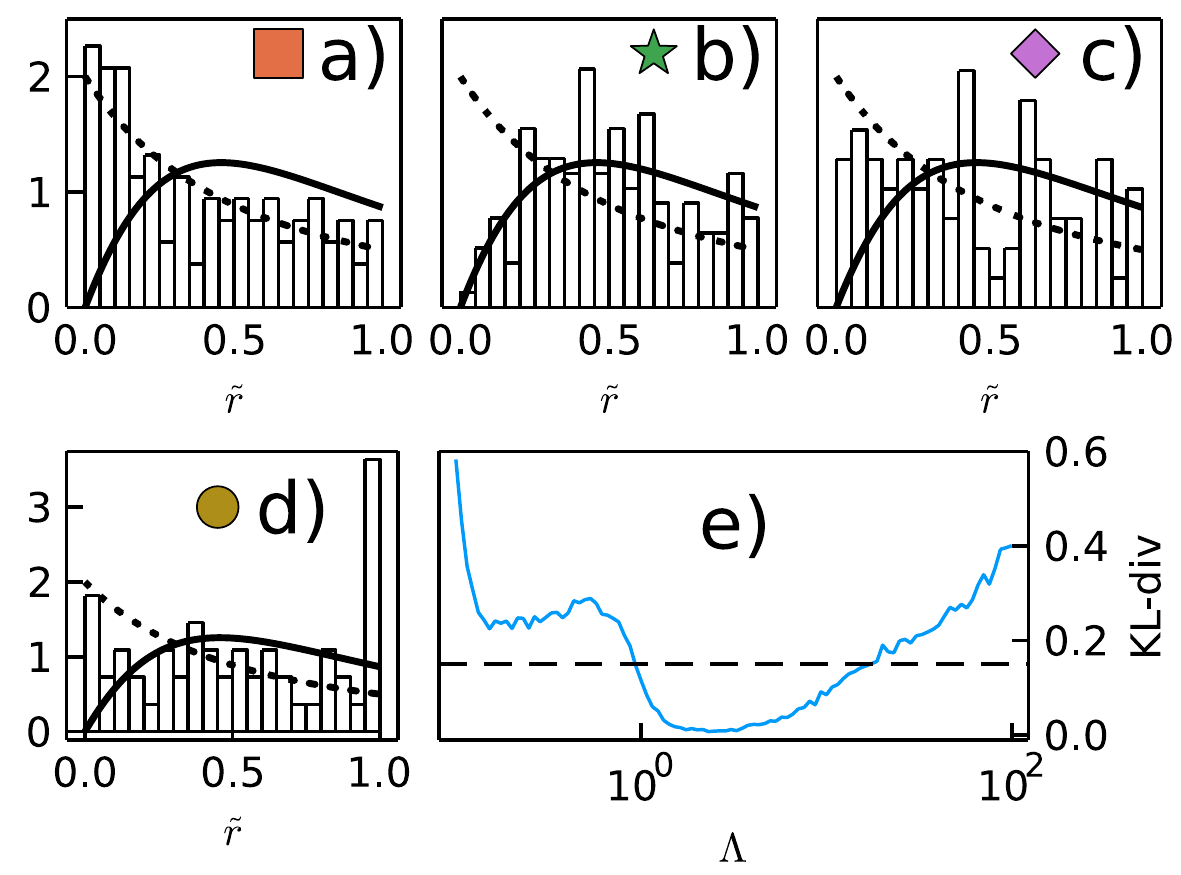}
\caption{\label{fig:rstat} \textbf{a)--d)}  Level ratio distributions for combinations
  of interaction  $\Lambda$ and relative energy $\tilde{E}$.
  \textbf{a)} $\Lambda\approx 0.28$ and $\tilde{E}=0.25$, \textbf{b)}
  $\Lambda\approx 2.48$ and $\tilde{E}=0.4$, \textbf{c)}
  $\Lambda\approx 12.33$ and $\tilde{E}=0.65$ and \textbf{d)}
  $\Lambda\approx 0.28$ and $\tilde{E}=0.13$.  Solid and dashed lines
  are $p_{\text{GOE}}(\tilde{r})$ and $p_{\text{Poi}}(\tilde{r})$,
  respectively.  \textbf{e)} KL divergence of the
  distribution of the level ratios over the full spectrum from the GOE distribution.
} 
\end{center}      
\end{figure}

In figures \ref{fig:rstat} a)-c) we show the observed ratio
distributions for three different combinations of relative energy
$\tilde{E}$ and interaction parameter $\Lambda$.  Since these
distributions are estimated from a finite number of energy eigenvalues
within the respective energy windows, they are shown as histograms.
The data here is extracted from calculations with $N=150$ bosons; 
the histograms are expected to converge to a smooth distribution in
the limit $N\to\infty$.

For visual guidance, the parameters $(\Lambda,\tilde{E})$ corresponding to
the panels in figure \ref{fig:rstat} are marked with respective
symbols in figure \ref{fig:chaos2d_rstat_mean}a).

The distribution in panel \ref{fig:rstat}a) is visually seen to be
close to the Poisson case.  Hence we expect the KL divergence from the
Poisson distribution ($D^{\text{Poi}}$) to be small and
the KL divergence from the GOE ($D^{\text{GOE}}$) to be
large.  The situation in panel \ref{fig:rstat}b) is the opposite
(close to GOE), while panel \ref{fig:rstat}c) shows an intermediate
case.  These expectations are borne out by the calculated KL
divergences:
\begin{align*}
\text{a)} \qquad & D^{\text{Poi}}\approx0.05, \quad   D^{\text{GOE}}\approx0.4; 
\\
\text{b)} \qquad & D^{\text{Poi}}\approx0.22, \quad   D^{\text{GOE}}\approx0.06; 
\\
\text{c)} \qquad & D^{\text{Poi}}\approx0.16, \quad   D^{\text{GOE}}\approx0.29. 
\end{align*}

In figure \ref{fig:chaos_2d}b), we used $D^{\text{GOE}}$ as a quantifier of chaos and presented its
values in the entire $(\Lambda,\tilde{E})$ plane as a heatmap.  We capped the values at $0.15$,
meaning that values $D^{\text{GOE}}>0.15$ are considered fully non-chaotic and are not
distinguished.  There is some arbitrariness in the exact choice of this value, but the main results
--- the overall shape in figure \ref{fig:chaos_2d}b) and its close correspondence with the classical
case, figure \ref{fig:chaos_2d}a) --- are not strongly affected by the use of a cutoff.  In figure
\ref{fig:rstat}e), we show $D^{\text{GOE}}$ for the complete energy spectrum as a function of
$\Lambda$, to provide a visual sense of the role of the cutoff in separating chaotic from
non-chaotic parameter values.

\subsection{Average ratio}

\begin{figure}
\begin{center}
\includegraphics[width=\columnwidth]{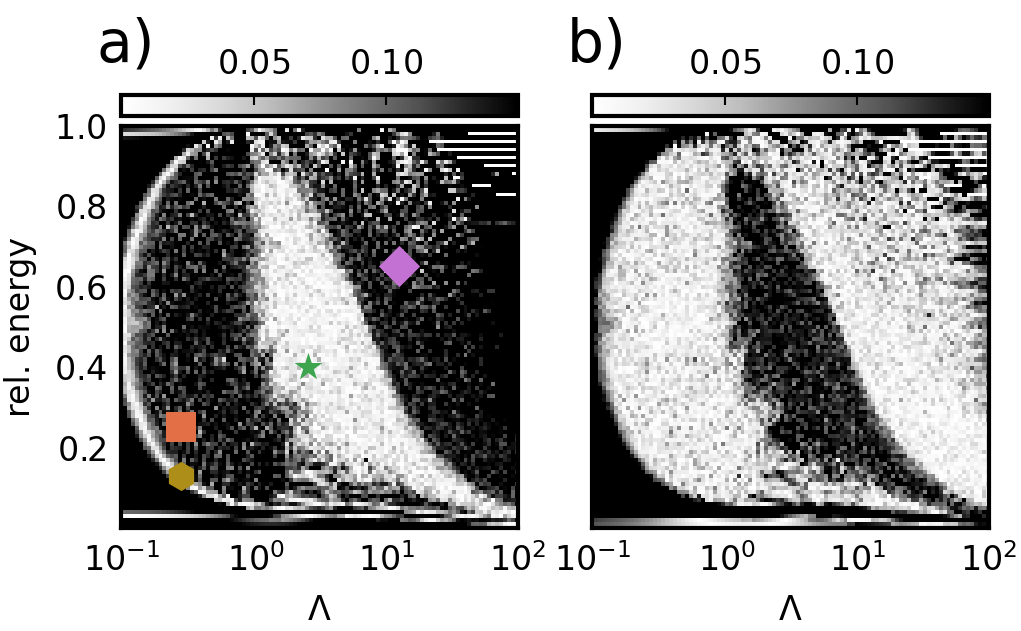}
\caption{\label{fig:chaos2d_rstat_mean} Distance of the mean of the
  level ratios $\langle \tilde{r} \rangle$ from a) the GOE level ratio
  mean $\langle \tilde{r} \rangle_{\rm GOE}$ and b)  the Poisson level
  ratio mean $\langle \tilde{r} \rangle_{\rm Poisson}$. The markers in a) indicate the systems shown in figure \ref{fig:rstat}. The square corresponds to figure \ref{fig:rstat} a), the star to b), the rhombus to c) and the circle to d).}
\end{center}
\end{figure}

Often only the first moment (mean) of the level ratio distribution is used as
a measure of  closeness to GOE or Poisson behavior. 

The mean of level ratios for the GOE and Poisson cases  are $\langle
\tilde{r} \rangle_{\rm GOE} \approx 4-2\sqrt{3} \approx 0.53590$ and
$\langle \tilde{r} \rangle_{\rm Poi} = 2\log 2 -1 \approx
0.38629$.  In cases a), b) and c)
above, the means are  0.39, 0.51, and 0.44.  i.e., they are
respectively close to $\langle\tilde{r}\rangle_{\rm Poi}$, close to
$\langle\tilde{r}\rangle_{\rm GOE}$, and intermediate, as expected. 

In figure \ref{fig:chaos2d_rstat_mean} we use the absolute distance from a)
$\langle\tilde{r}\rangle_{\rm GOE}$ or from b) $\langle\tilde{r}\rangle_{\rm Poi}$ as possible
alternate measures of chaos.  Compared to Figure \ref{fig:chaos_2d} b), we see that the same
information is captured --- a more chaotic region at intermediate $\Lambda$ and intermediate
$\tilde{E}$ is clearly visible in both these cases.  Overall, the mean of level ratios is closer to
$\langle\tilde{r}\rangle_{\rm GOE}$ inside this region and closer to
$\langle\tilde{r}\rangle_{\rm Poi}$ outside.  Even the fine structures at the boundary between the
two regions, previously seen in the classical case in figure \ref{fig:chaos_2d}a), are visible.

However, there are some artifacts.  The most prominent is the arc at the left (small $\Lambda$)
region, in figure \ref{fig:chaos2d_rstat_mean}a) The reason is that, at small $\Lambda$, the
spectrum shows features specific to the free-boson case, deviating from the Poisson model of
completely uncorrelated models. We can see this in figure \ref{fig:rstat}d), which corresponds to a
$(\Lambda,\tilde{E})$ combination falling on the arc of panel \ref{fig:chaos2d_rstat_mean}a).  The
distribution is neither Poisson-like nor GOE-like: it is nonzero for $\tilde{r}\to0$ and has a
pronounced peak at $\tilde{r}\to1$.  Together, these lead to an average
$\langle\tilde{r}\rangle\approx0.52$ which is coincidentally close to
$\langle\tilde{r}\rangle_{\rm GOE}$. The deviation from Poisson at very small $\Lambda$ is also seen
in panel \ref{fig:chaos2d_rstat_mean}b), in the form of a darker region at the very left of the
heatmap.

To summarize: the chaos-regular demarcation in the
$(\Lambda,\tilde{E})$ plane can also be visualized using the mean
$\langle\tilde{r}\rangle$, modulo some artifacts.

\section{Eigenstate statistics}\label{sec:eigenstates}

In this section, we discuss characterizing chaoticity in the Bose-Hubbard system using the deviation
of eigenstate structure from those of GOE matrices.  We describe the calculations leading to figure
\ref{fig:chaos_2d}c).

The eigenstates of random GOE matrices are uniformly distributed on the $(D-1)$ dimensional unit
sphere $S^{D-1}$.  For large $D$, a uniform distribution on $S^{D-1}$ is well approximated by a
$D$-dimensional Gaussian distribution with independent entries and mean zero and variance $1/D$
\cite{Pinelis_ESAIM2015_Maths, Pinelis_EJournStat2016_Maths}.
To compare the Bose-Hubbard system with the GOE case, we will compare the coefficients of
Bose-Hubbard eigenstates with the Gaussian distribution with mean 0 and variance $1/D$.  It has been
observed that mid-spectrum eigenstates of chaotic/ergodic many-body systems generally have
coefficients with a near-Gaussian distribution \cite{Luitz_BarLev_PRL2016_anomalous,
  Beugeling_Baecker_Haque_PRE2018, Khaymovich_Haque_McClarty_PRL2019,
  Baecker_Haque_Khaymovich_PRE2019, Luitz_Khaymovich_BarLev_multifrac_SciPost2020,
  DeTomasi_Khaymovich_Pollmann_Warzel_PRB2021, Prosen_Sotiriadis_PRL2021, Ueda_LocalRMT_PRL2021,
  Pausch_Buchleitner_PRL2021_BHchaos, Haque_McClarty_Khaymovich_PRE2022_entanglement}, in accord
with Berry's conjecture \cite{Berry_JPA1977}.  On the other hand, eigenstates of integrable or
many-body-localized systems, as well as eigenstates at the spectral edges of nominally chaotic
systems, typically have markedly non-Gaussian distributions \cite{DeLuca_Scardicchio_EPL2013,
  Luitz_BarLev_PRL2016_anomalous, Beugeling_Baecker_Haque_PRE2018,
  Luitz_Khaymovich_BarLev_multifrac_SciPost2020}.

To compare distributions of eigenstate coefficients, we used the excess kurtosis, $\kappa$, of the
set of coefficients.  The kurtosis is the fourth standardized moment. The excess kurtosis of a
distribution is the difference between the kurtosis of that distribution and the kurtosis of a
Gaussian distribution, which is 3.  Thus, large values of $\kappa$ represent larger deviations from
Gaussianity and hence from GOE/chaotic behavior, whereas small values represent more chaotic
behavior.  When we report values of the kurtosis, we always mean the excess kurtosis $\kappa$, even
when the word `excess' is omitted.

The deviation of many-body eigenstates from Gaussianity could also be measured using the KL
divergence, as in \cite{Beugeling_Baecker_Haque_PRE2018}, or using the inverse participation ratio
(IPR) or multifractal exponents, as in \cite{DeLuca_Scardicchio_EPL2013,
  Beugeling_entanglement_JSM2015, Baecker_Haque_Khaymovich_PRE2019,
  Luitz_Khaymovich_BarLev_multifrac_SciPost2020, DeTomasi_Khaymovich_PRL2020,
  Pausch_Buchleitner_PRL2021_BHchaos, Pausch_Buchleitner_NJP2021_BHchaos}.  We expect these
measures to provide very similar overall pictures as the one we present using the kurtosis.  In
fact, when the mean is negligible (which is true for most eigenstates excepting some at the spectral
edges), the kurtosis is proportional to the IPR.

Eigenstate coefficients are defined with respect to a basis.
We will investigate eigenstates of the Bose-Hubbard system with respect to three bases:
\begin{enumerate}
\item the computational basis, which is given by the mutual eigenstates of the number operators
  $n_i$;
\item the mutual eigenbasis of the hopping operators $a_i^\dagger a_j$, i.e., the eigenbasis of the
  non-interacting (free) system;
\item the eigenbasis of a perturbed free system with hopping terms $J_{ij}=1$ and small on-site
  perturbing potentials $\sum_i\epsilon_{i}n_{i}$ with values $\epsilon_1=-0.01$, $\epsilon_2=0.02$
  and $\epsilon_3=-0.03$ on the three sites.
\end{enumerate}
We name the kurtosis of the coefficients in the three bases as $\kappa_1$,  $\kappa_2$,  $\kappa_3$
respectively.  
In Figure \ref{fig:chaos_2d}c), the quantity presented is obtained from a combination of the first
and third choices above, namely $\max(\kappa_1,\kappa_3)$. 

We assume that the distributions underlying the eigenstate components
of two eigenstates close in energy are similar. As before, we divide
the energy spectrum of each Bose-Hubbard system with interaction
strength $\Lambda$ into 100 equally spaced intervals and refer to them
by their relative energy $\tilde{E}$. We compute the kurtosis $\kappa$
for every eigenstate and average the calculated kurtosis over each
energy interval.  If the mean is zero, the averaged kurtosis in an
energy interval equals the kurtosis of all components of all
eigenstates in that energy interval.

\begin{figure}
\begin{center}
\includegraphics[width=\columnwidth]{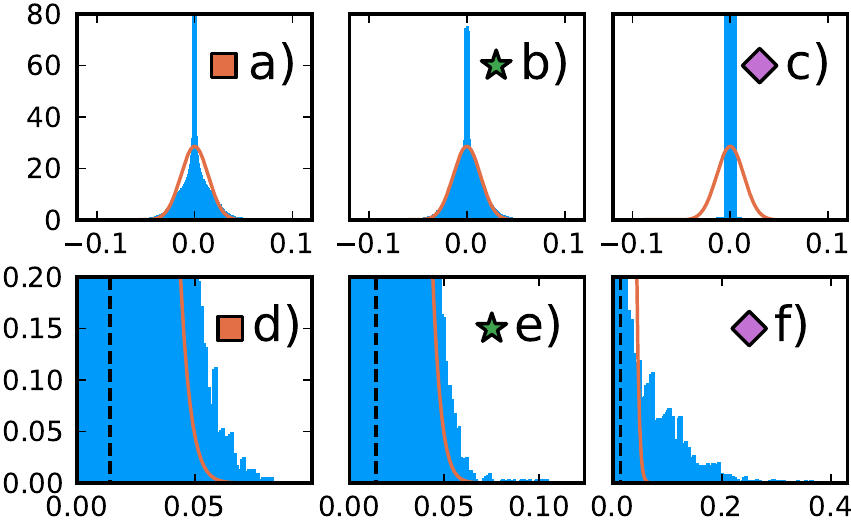}
\caption{\label{fig:eigenstate_density} Histograms of eigenstate components of Bose-Hubbard systems
  with $N=100$ particles, in the computational basis.  In \textbf{a)}-\textbf{c)}, the interaction
  and energy intervals, $(\Lambda,\tilde{E})$, are the same as those used in figure
  \ref{fig:rstat}\textbf{a)}-\textbf{c)}.  Panels \textbf{d)}-\textbf{f)} are zoomed into the right
  tails of \textbf{a)}-\textbf{c)}. The dashed line indicates the standard deviation. The orange
  line is a Gaussian density with mean 0 and standard deviation $1/D$. The excess kurtosis is
  \textbf{a)} $\kappa_1\approx 1.6$, \textbf{b)} $\kappa_1\approx 0.8$, \textbf{c)} $\kappa_1\approx 122$.
}
\end{center}
\end{figure}

\begin{figure*}
\begin{center}
\includegraphics[width=\textwidth]{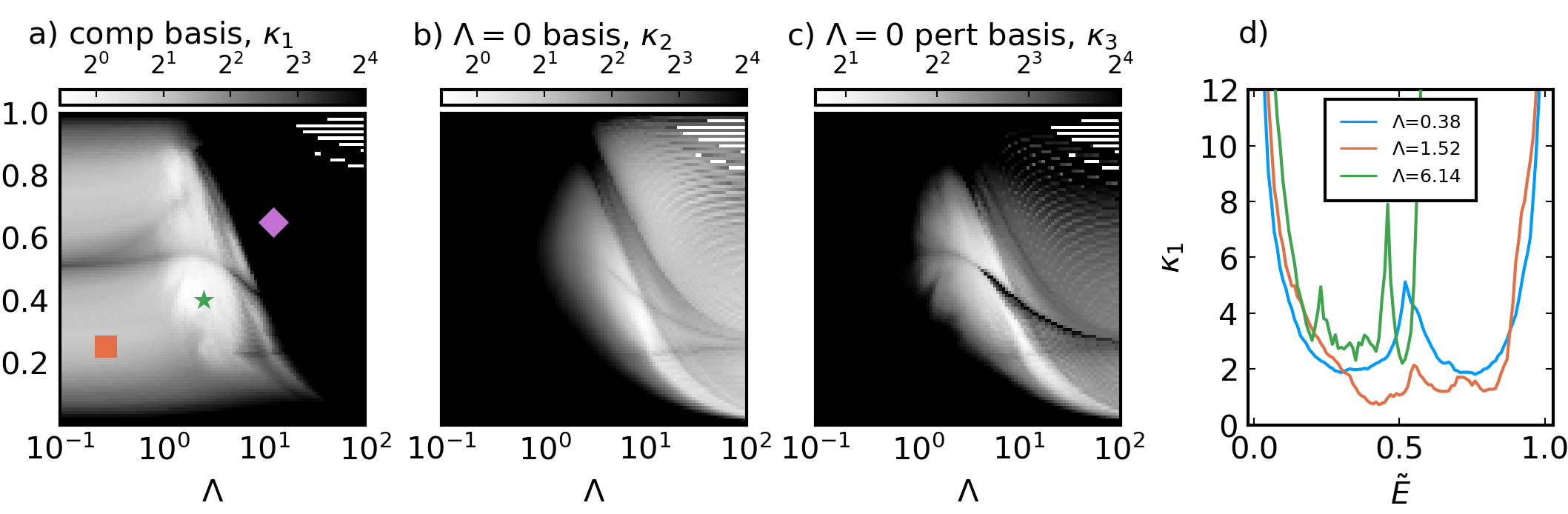}
\caption{\label{fig:chaos2d_kurtosis} Heatmaps of kurtosis of eigenstate coefficients, for three
  different bases, listed in the text.  The kurtosis is cut off at $2^4$ in each case.    Markers in
  a) indicate the $(\Lambda,\tilde{E})$ values for which histograms are shown in Figure
  \ref{fig:eigenstate_density}. 
  %
  % , with square, star, rhombus corresponding to a), b), c) respectively.
  %
  The kurtosis heatmap shown in the Introduction, Figure \ref{fig:chaos_2d}b), is a combination of panels a) and
  c) here --- for each  $(\Lambda,\tilde{E})$, the larger of the two values is chosen in Figure
  \ref{fig:chaos_2d}b). 
}
\end{center}
\end{figure*}

In Figure \ref{fig:eigenstate_density}a)-c), we show the eigenstate coefficient distributions in the
computational ($n_i$) basis, for the three $(\Lambda,\tilde{E})$ combinations used previously in
Figure \ref{fig:rstat}.  (Visual guidance to these three parameter combinations is provided in
Figure \ref{fig:chaos2d_rstat_mean}a) and \ref{fig:chaos2d_kurtosis}a) using corresponding symbols.)
The calculated excess kurtosis for these cases are respectively $\kappa_1\approx1.6$,
$\kappa_1\approx 0.8$ and $\kappa_1\approx 122$.  The case b) is thus closest to Gaussian, followed
by a), while case c) is very far from Gaussian.  This is consistent with the visual appearance of
the full distributions. It is also consistent with the comparison of the tails of the distibutions
against the tails of the Gaussian distribution, as shown in d)-f).

We note in Figure \ref{fig:eigenstate_density}b) that the coefficient distribution, although closest
to Gaussian, has a large peak near zero.  Even in the most chaotic region of the
$(\Lambda,\tilde{E})$ plane, the eigenstates depart considerably from the random-matrix case.  This
is a manifestation of the three-site Bose-Hubbard system being a very mixed system.

In Figure \ref{fig:chaos2d_kurtosis}a) we show the kurtosis $\kappa_1$ for eigenstates in the
computational basis as a heatmap in the $(\Lambda,\tilde{E})$ plane.  Comparing with previous
sections, we see that small $\kappa_1$ correlates with non-zero Lyapunov exponents and GOE level
statistics, while intermediate and large $\kappa_1$ correlates with zero Lyapunov exponents and
non-GOE level statistics.  The shape of the small-$\kappa_1$ region matches the more chaotic region
identified previously using classical Lyapunov exponents or using level statistics.  Even subtle
features from the heatmaps in the previous sections, such as the bulges around $\Lambda\approx 1$
and $\tilde{E} \approx 0.5$ are visible.

For small $\Lambda$, the kurtosis in the computational basis in Figure \ref{fig:chaos2d_kurtosis}a)
shows intermediate rather than large kurtosis, thus failing to fully capture the highly non-chaotic
nature of the system in this region.  The reason is probably that the small-$\Lambda$ eigenstates
are so different from the computational basis states (which are $\Lambda\to\infty$ eigenstates) that
they have overlap with a large number of the basis states, leading to a small IPR (hence small
kurtosis).  

A complementary view is obtained via $\kappa_2$ in figure \ref{fig:chaos2d_kurtosis}b), where
$\Lambda=0$ eigenstates have been used as basis.  (Because of the large degeneracy at $\Lambda=0$,
there is some computational arbitrariness in the choice of this basis.)  This basis now has the
opposite problem --- it fails to show the non-chaotic nature of large-$\Lambda$ region.  The problem
is partially alleviated by choosing as basis the eigenstates of a non-interacting Hamiltonian with
small on-site perturbing potentials; the resulting excess kurtosis $\kappa_3$ is shown in figure
\ref{fig:chaos2d_kurtosis}c).

For random-matrix eigenstates, one expects Gaussian behavior with respect to almost any basis.  In
figure \ref{fig:chaos2d_kurtosis}, the high-chaos region is marked by low kurtosis in all three
basis choices, consistent with the idea of basis-independence.  The other regions appear more or
less Gaussian-like depending on basis choice.  To demarcate the highly chaotic region from less
chaotic regions, we can use a combination of kurtosis calculations, taking the larger one from the
kurtosis obtained in the first and third basis, i.e., $\max(\kappa_1, \kappa_3)$.  This is what we
did in figure \ref{fig:chaos_2d}b).

\section{Scaling  of eigenstate expectation values}\label{sec:eth}

According to the Eigenstate Thermalization Hypothesis (ETH) \cite{Deutsch_PRA1991,
  Srednicki_PRE1994, Srednicki_1999, rigol_thermalization_2008}, the expectation values of physical
operators in eigenstates become smooth functions of energy in the thermodynamic limit --- the
fluctuations of eigenstate expectation values (EEV's) decreases with system size in a specific
manner \cite{Beugeling_Moessner_Haque_PRE2014}.  Instead of the thermodynamic limit, one could also
ask how EEV fluctuations die off in the classical limit $N\to\infty$ \cite{Nakerst_Haque_PRE2021},
which is the limit we are considering in this work.
We examine the operator
\[
\bar{A} = \frac{a_2^\dagger a_1}{N}.  
\]
The fluctuations of the EEV's of this operator, $\sigma(\bar{A})$, scales as a power law,  
\[
  \sigma(\bar{A}) \sim D^{-e} ,
\]
with the Hilbert space dimension $D=\binom{N+L-1}{L-1}$.  If the eigenstates of the system were
fully chaotic, i.e., if the eigenstate coefficients were well-approximated by Gaussian independent
variables, one can show that the scaling exponent would be $e=\frac{1}{2}$
\cite{Nakerst_Haque_PRE2021}.  The deviation from $e=\frac{1}{2}$ is thus a measure of departure
from chaoticity.

In Figure \ref{fig:chaos_2d}d), we have presented a heatmap of the exponents $e$, determined
numerically, for each energy window and interaction parameter.  The exponents are determined by
fitting the $\sigma(\bar{A})$ vs $D$ data, for system sizes ranging from $N=90$ to $N=170$ in steps
of $10$, i.e., $D$ ranging from $\approx4000$ to $\approx15000$.  As found in
Ref.~\cite{Nakerst_Haque_PRE2021}, even in the most chaotic regions of the $(\Lambda,\tilde{E})$
plane, the exponent falls well below the ideal value $\frac{1}{2}$.  The numerically observed
exponent ranges from $0$ in the regular regions to $\approx0.3$ in the most chaotic regions.  The
resulting heatmap, Figure \ref{fig:chaos_2d}d), is noisier and less sharp than those obtained from
the other measures discussed in previous sections.  But the overall demarcation of chaotic and
non-chaotic regimes is clearly visible.

\section{Context and Discussion}\label{sec:discussion}

In this article we considered a quantum many-body model that has a classical limit and is well-known
to be `mixed', the Bose-Hubbard trimer.  We compared the classical Lyapunov exponents of the
classical limit against quantum measures of chaos obtained from eigenvalues (statistics of level
spacing ratios) and eigenstates (coefficient statistics, fluctuations of expectation values of an
operator).  Overall, the agreement in the chaos-regular demarcation between the classical case and
the various quantum measures is very good.  

This reflects the general agreement of chaos measures between quantum systems and their classical
limit, when such a limit exists, observed computationally in many different Hamiltonian systems over
decades.  Perhaps most prominently, single-particle systems in a potential (`billiards') have an
obvious classical limit --- one simply treats the system classically.  The literature on
quantum-classical correspondence in different billiard systems is vast.  Other than billiards,
classical and quantum chaos measures have been compared in coupled rotors or coupled tops
\cite{Feingold_Peres_PhysicaD1983_coupled_rotators, Feingold_Moiseyev_Peres_PRA1984,
  Ballentine_PRA2004, Fan_Gnutzmann_Liang_PRE2017_FeingoldPeres,
  Mondal_Sinha_PRE2020_twocoupledtops},

bosonic systems
\cite{Weiss_Teichmann_PRL2008_BHdimer_quantumclassical, Kolovsky_IJMPB2016, Rautenberg_Gaertner_PRA2020}, 
the Dicke model and other spin-boson systems
\cite{Hirsch_PRA2014_Dicke_comparative2, Hirsch_PhysicaScrpta2015_Dicke,
  Ghosh_Sinha_PRE2016_kicked_Dicke, Hirsch_PRE2016_Dicke, 
  Carlos_Santos_Hirsch_PRL2019_LyapunovOTOCDickemodel, Bollinger_AMRey_NatureComm2019_Dicke,
  Pappalardi_PRA2020_semiclassicalsystems_kickedtop_dickemodel, Robnik_PRE2020_Dicke,
  LSantos_Hirsch_NJP2020_Dicke, SantosHirsch_NJP2021_spinbosonscarring},
the Sherrington-Kirkpatrick model 
\cite{Pappalardi_Polkovnikov_Silva_SciPost2020_SherringtonKirkpatrick}, 
and spin systems
\cite{Robb_Reichl_PRE1998_twospin, Emerson_Ballentine_PRA2001_twospin,
  Pappalardi_Silva_Fazio_PRB2018_longrange, Ray_Sinha_Sen_PRE2019_QuasiperiodicallyDrivenSpin}. 
%
% and various other models \cite{Porter_Barr_Reichl_PRE2017_Sinailattice}.
%
A common theme is that, for spin systems or systems
with angular momentum, the large-spin (or large angular momentum) limit is the classical limit,
whereas for bosonic systems, the large-boson-number system is the classical limit.

For classical systems, it is common in the literature to demarcate chaotic and non-chaotic regimes
using Poincar\'e sections \cite{Feingold_Peres_PhysicaD1983_coupled_rotators,
  Bohigas_Tomsovic_Ullmo_PhysRep1993, Dullin_Richter_Wittek_Chaos1996, Robb_Reichl_PRE1998_twospin,
  Robb_Reichl_PRE1998_twospin, Hirsch_PRA2014_Dicke_comparative2,
  Hirsch_PhysicaScrpta2015_Dicke,Robnik_PRE2020_Dicke, LSantos_Hirsch_NJP2020_Dicke}.  We have
focused instead on the Lyapunov exponent, and presented it as a multi-branched function of energy.
Inspired by Ref.\ \cite{Carlos_Santos_Hirsch_PRL2019_LyapunovOTOCDickemodel}, we have used the
fraction of Lyapunov exponents which are positive as a chaos measure, and compared it with other
ways of exploiting the FTLE results to demarcate highly chaotic and less chaotic behaviors.  It is
clear that, if the phase space at fixed energy is separated into regular and chaotic regions, then
the Lyapunov exponent plotted against energy (with many phase space points sampled in each energy
window) will have to be a multi-valued plot.  We hope that explicitly presenting and analyzing this
multi-valued dependence will contribute to the intuition available on mixed systems.

For quantum systems, we used several measures: (1) the statistics of level spacing ratios based on
eigenvalues alone; (2) the coefficients of eigenstates, based on eigenstates expressed in different
bases; (3) the scaling of fluctuations of eigenstate expectation values (EEVs), based on eigenstate
properties.  Level spacing statistics and eigenstate coefficients have been been considered and
used as chaos measures for several decades.  The EEV fluctuation scaling is based on understanding
that has emerged in recent years, motivated by studies of thermalization and ETH.  

Of course, there are other interesting measures of quantum chaos that could be considered for
comparison.  A new candidate is the out-of-time-ordered correlator or OTOC
\cite{MaldacenaShenkerStanford_JHEP2016, Stanford_JEHP_2016,
  Rammensee_Urbina_Richter_PRL2018_semiclassicOTOCs,
  Carlos_Santos_Hirsch_PRL2019_LyapunovOTOCDickemodel} whose behavior (exponential growth) defines a
quantum Lyapunov exponent for chaotic systems.  For our mixed system, we were unable to
unambiguously identify or rule out exponential regimes in the dynamics, for the parameter
combinations we attempted.  It remains unclear to us whether the OTOC is a useful measure for
numerically demarcating more-chaotic parameter regimes from less-chaotic and non-chaotic parameter
regimes in mixed systems.

There are some peculiar features in both the eigenvalue and eigenstate statistics, whose origins
remain unclear and might be clarified in future studies.  In figure \ref{fig:chaos2d_rstat_mean}a,
the arc in the small-$\Lambda$ part of the heatmap is due to the level spacing having peculiar
statistics, as shown in figure \ref{fig:rstat}d), due to a significant number of successive equal
spacings.  In the eigenvector statistics, there are some mid-spectrum states that are highly
non-Gaussian, even at intermediate $\Lambda$, as seen through the dark nearly horizontal line in
figure \ref{fig:chaos2d_kurtosis}a) at intermediate energies, and the dark curved line in figure
\ref{fig:chaos_2d}c), running through the more-chaotic light-colored region at intermediate
energies.  These might be due to scar-like eigenstates embedded within the more-chaotic eigenstates.
Presumably, such peculiar features are less likely to appear in more fully ergodic systems, such as
the Bose-Hubbard system with larger number of sites.

\begin{acknowledgments} 

The authors thank Rodrigo Camara, Quirin Hummel, Andrey Kolovsky, Rubem Mondaini, Pedro
Ribeiro, and Lea Santos for helpful discussions. GN gratefully acknowledges support from the Irish
Research Council Government of Ireland Postgraduate Scholarship Scheme (GOIPG/2019/58) and the
Deutsche Forschungsgemeinschaft through SFB 1143 (project-id 247310070). Part of the calculations
were performed on the Irish Center for High-End Computing (ICHEC).  

\end{acknowledgments}

\appendix

\section{Lyapunov exponents for more chaotic systems}\label{sec:appendix_FTLE_larger_systems}

\begin{figure}
\begin{center}
\includegraphics[width=\columnwidth]{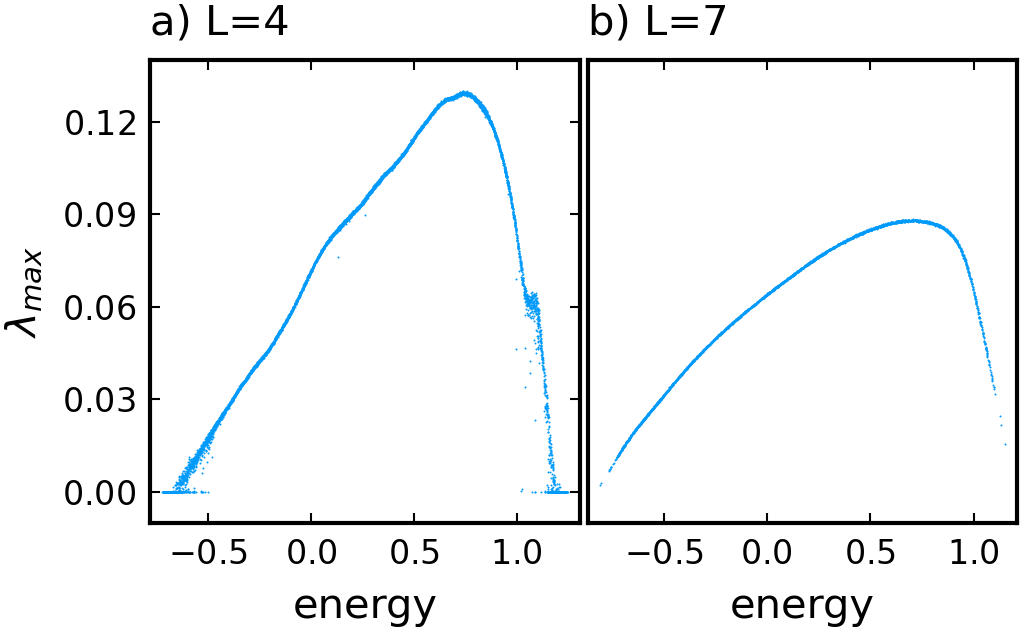}
\caption{\label{fig:lyapunov_vs_energy_large_L} FTLE estimates for the classical largest Lyapunov
  exponent $\lambda_{\text{max}}$ plotted against energy, for the classical limits of Bose-Hubbard
  chains with $L=4$ (left) and $L=7$ (right) sites.  The interaction parameter is $\Lambda=1.52$. The
  variational equations were evolved to time $t=10^6$.}
\end{center}
\end{figure}

In figure \ref{fig:lyapunov_vs_energy_large_L} we present, for comparison, FTLE's calculated for the
4-site chain and the 7-site chain.  The classical Hamiltonian is the same as that presented in
Section \ref{sec:model}.   

The systems are increasingly more chaotic with increasing $L$.  The arguably most remarkable
signature of mixedness in the $L=3$ case was the multi-branched behavior of the Lyapunov exponents, as
presented in figure \ref{fig:lyapunov_vs_energy} in the main text.  For the $L=4$ case, which is
more chaotic, some signature of the same phenomenon can be seen at small and large energies, figure
\ref{fig:lyapunov_vs_energy_large_L}a).  In the $L=7$ case, which is much more chaotic, the
phenomenon is absent, figure \ref{fig:lyapunov_vs_energy_large_L}b).  

Obtaining good estimates for the Lyapunov exponents is more challenging for mixed systems.
Comparing figure \ref{fig:lyapunov_vs_energy} and the two panels of figure
\ref{fig:lyapunov_vs_energy_large_L}, we see cleaner (less noisy) data for larger $L$, for the same
time of propagation, even though there are more variables/equations ($2L$) to be evolved for larger
$L$.

\section{Numerical calculation of Lyapunov exponents}\label{sec:appendix_numerical_Lyapunov_exponents}

In this Appendix, we provide an outline of how the largest Lyapunov exponent are estimated
numerically \cite{Benettin_Mec1980_lyapunov1, Benettin_Mec1980_lyapunov2}.  Calculation of the full
set of Lyapunov exponents is slightly more involved and so not described here.

We will use the symbol $\psi(t)$ to represent a state or phase space point of the classical system
at time $t$.  For the Bose-Hubbard chain, $\psi(t)$ is a vector of $L$ complex variables, or $2L$
real variables.  The symbol $\psi_0$ will be used for the initial state, i.e., $\psi_0=\psi(0)$.
(In the main text, subscripts to $\psi$ have been used as site indices, but there should be no
confusion with our use of the subscript $0$ here.)

Intuitively, the largest Lyapunov exponent $\lambda_{\rm max}$ is given by
\begin{equation}\label{eq:def_lambda_max_1_appendix}
	e^{t \lambda_{\rm max}} \approx \frac{ \| \tilde{\psi}(t) - \psi(t) \| }{ \| \tilde{\psi}_0 - \psi_0\| },
\end{equation}
at large times.  Here $\psi(0)=\psi_0$ and $\tilde{\psi}(0)=\tilde{\psi}_0$ are two initial states which are `close'
to each other, and $\psi$ and $\tilde{\psi}$ are time-evolving according to Hamilton's equations of
motion $i\frac{d}{dt}\psi = \frac{\partial \mathcal{H}}{\partial \psi^*}$. The largest Lyapunov
exponent $\lambda_{\text{max}}$ is independent of the choice of the norm in
eq.~\eqref{eq:def_lambda_max_1_appendix}, as long as the phase space is finite-dimensional.

Eq.~\eqref{eq:def_lambda_max_1_appendix} implies that if the largest Lyapunov exponent
$\lambda_{\text{max}}$ is positive the two states $\psi$ and $\psi'$ separate exponentially, while a
zero largest Lyapunov exponent $\lambda_{\text{max}}=0$ means an at most polynomial spread. For
Hamiltonian systems the Lyapunov exponents come in pairs of equal magnitude and opposite sign, so
that the largest Lyapunov exponent $\lambda_{\text{max}}$ is at least 0.  This is a consequence of
Liouville's theorem.

To calculate the Lyapunov exponent numerically, one might be tempted to choose two initially close
states and calculate the right hand side of eq.~\eqref{eq:def_lambda_max_1_appendix} for large times
$t$.  Unfortunately, this does not work for bounded systems.

This problem is circumvented by solving the so called variational equations. Denote the time-evolution of the dynamical system corresponding to Hamilton's equation of motion as
\begin{equation}
	\Phi(t, \psi_0) = \psi(t), \; \text{where } \psi(0) = \psi_0.
\end{equation}
The dynamical system obeys the semi-group property $\Phi(t+s, x_0) = \Phi(s, \Phi(t, x_0))$ for all times $t$ and $s$. Eq.~\eqref{eq:def_lambda_max_1_appendix} now reads in terms of $\Phi$ and $\phi_0 = \psi_0 - \tilde{\psi}_0$ as
\begin{equation}\label{eq:def_lambda_max_1_Phi}
	e^{t\lambda_{\text{max}}} \approx \frac{\| \Phi(t, \psi_0 + \phi_0) - \Phi(t, \psi_0) \|}{ \| \phi_0 \|}.
\end{equation}
By linearizing Eq.~\ref{eq:def_lambda_max_1_Phi} we obtain the largest Lyapunov exponent as
\begin{equation}\label{eq:def_lambda_max_2_appendix}
\lambda_{\text{max}} = \lim_{t\to\infty} \frac{1}{t} \log \left \| \partial_\psi\Phi(t, \psi)|_{\psi=\psi_0} \cdot \frac{\phi_0 }{\|\phi_0\|} \right\|.
\end{equation}
Note that $\partial_\psi \Phi$ is in general a matrix so the product $\cdot$ denotes the
matrix-vector product. The existence of the above limit is ensured by Osedelets theorem
\cite{Oseledets_MosMS1968}.  One can show that $\partial_\psi \Phi$ evolves in time according to so called variational equations
\begin{equation}\label{eq:def_var_eqs_appendix}
	i \frac{d}{dt} \partial_\psi \Phi(t, \psi) = \partial_\psi \partial_{\psi^*} \mathcal{H}(\psi(t)) \cdot \partial_\psi \Phi(t, \psi),
\end{equation}
where $\partial_\psi \partial_{\psi^*} \mathcal{H}$ denotes the Hessian of the Hamiltonian $\mathcal{H}$ in the variables $\psi$ and $\psi^*$ and the initial condition is $\partial_\psi\Phi(0, \psi_0)=\operatorname{Id}$. 

In eq.~\eqref{eq:def_lambda_max_2_appendix} the knowledge of the full matrix $\partial_\psi\Phi$ is not required. Only the deviation vector $\phi(t) = \partial_\psi\Phi(t, \psi)|_{\psi=\psi_0} \phi_0 / \|\phi_0\|$ is needed. The deviation vector evolves according to the variatonal equations \eqref{eq:def_var_eqs_appendix} as well. This follows from the linearity of $d/dt$.

In principle one could now evolve Hamilton's equations together with the variational equations to
obtain $\phi(t)$ for some large time $t$ and determine $\lambda_{\rm max}$ via
eq.~\eqref{eq:def_lambda_max_2_appendix}.  However, for positive $\lambda_{\rm max}$ the norm of
$\phi(t)$ will blow up exponentially and will quickly be unmanageable by finite precision. This is
circumvented by renormalizing $\phi(t)$ and restarting the time evolution, whenever it becomes too
large. 

\bibliography{lit}

\end{document}